\documentclass[
 amsmath,amssymb,
 aps,
]{revtex4-2}
\usepackage[T1]{fontenc}
\usepackage{graphicx}% Include figure files
\usepackage{dcolumn}% Align table columns on decimal point
\usepackage{bm}% bold math
\usepackage[separate-uncertainty=true,separate-uncertainty-units=bracket]{siunitx}
\DeclareSIUnit{\dBm}{dBm}

\usepackage[breaklinks]{hyperref}

\usepackage{xcolor}
\usepackage{ulem} %\sout
\usepackage{caption}
\usepackage{subcaption}
\usepackage{overpic}
\usepackage{lpic}
\newcommand{\Figure}[1]{Figure~\ref{fig:#1}}
\newcommand{\Equation}[1]{Equation~\ref{eq:#1}}

\newcommand{\loss}{51.34} %dB
\newcommand{\distance}{\SI{300.052}{\kilo\meter} } % km
\newcommand{\drouputDuration}{\SI{18}{\second} }
\newcommand{\uptime}{\SI{99.86}{\percent} }

\begin{document}

\title{
Unrepeated White Rabbit Time Synchronisation over a 300~km Optical Fibre Link 
}% Force line breaks with \\

\author{Ben Amies-King}
 \email{ben.amies-king@york.ac.uk}
\author{Marco Lucamarini}%
\affiliation{%
School of Physics, Engineering \& Technology and York Centre for Quantum Technologies\\
University of York, YO10 5FT York, U.K.
}%

\date{\today}% It is always \today, today,

\begin{abstract}
\noindent White Rabbit (WR) technology provides a commercially-available off-the-shelf solution for time synchronisation with sub-nanosecond accuracy and picosecond-level precision over optical fibre links typically spanning tens of kilometres.
Such high-performance time dissemination can support a variety of applications, including position, navigation and timing (PNT), financial transactions, metrology, as well as entanglement and quantum key distribution (QKD).
Demonstrations of WR over significantly longer distances remain few and far between, particularly in scenarios where intermediate amplification is unavailable, such as stretches of long-haul underwater fibre.
In this work, we report the longest unrepeated deployment of WR to date, achieving time synchronisation over a 300~km (\SI{\loss}{\deci\bel}) single-span optical fibre link, even in highly asymmetrical configurations, with \uptime uptime, whilst maintaining picosecond-level precision and sub-nanosecond accuracy. This was achieved through careful selection and optimisation of the components deployed at the link’s end points.
By leveraging standard telecom fibre and off-the-shelf hardware, our results pave the way for a scalable and standardised timing backbone for large-scale quantum networks, offering a practical route toward time distribution in future heterogeneous quantum communication systems.
\end{abstract}
\maketitle

\section{Introduction}
\noindent Time is one of the most fundamental physical quantities, underlying measurement, communication, and control in complex systems. In distributed networks, maintaining a common notion of time across all nodes is essential for accurate coordination and reliable operation. The White Rabbit (WR) technology \cite{lipinski_white_2011}
is a high-precision timing protocol that extends the Ethernet Precision Time Protocol (PTP) \cite{ieee_ptp_2010} to achieve sub-nanosecond accuracy and picosecond-level precision \cite{cern_wr-website_2024}.
WR combines the existing Ethernet network timing standards PTP and synchronous Ethernet (SyncE) with digital dual-mixer time difference (DDMTD) clock recovery to deliver deterministic timing performance far beyond conventional GPS-based systems.
Its open-source hardware and software design, along with wide commercial availability \cite{safran_wr-len_2024}, make WR an attractive solution for a broad range of time-critical applications, including large-scale scientific experiments \cite{loschmidt_white_2009, calvo_subnanosecond_2020}, telecommunications, and power grid management \cite{liu_enhancing_2025}.
In particular, its excellent timing precision is motivating increasing integration with timing equipment such as time-to-digital converters (TDCs) \cite{li_integration_2017}, along with experimental demonstrations of WR-supported distributed TDC networks \cite{pan_high_2014}.

Among the most promising emerging applications of WR are those related to the distribution of quantum keys \cite{bennett_quantum_2014} and entanglement \cite{ekert_quantum_1991,bennett_quantum_1992}, which enable information-theoretically secure communication and enhanced quantum processing capabilities. 
Quantum key distribution (QKD), in particular, offers provable security guaranteed by the laws of quantum mechanics, unlike classical cryptographic methods, which rely on assumptions about computational hardness and may become vulnerable to future computers.
As the only method of establishing a shared symmetric key that is fundamentally secure against both classical and quantum adversaries, QKD is expected to play a key role in protecting highly sensitive data pathways with financial, governmental, or national security implications.
Significant progress has already been made towards commercial deployment, with several vendors now offering QKD systems \cite{toshiba_qkd, idquantique_qkd, luxquanta_qkd, thinkquantum_qkd, qti_qkd, qbird_qkd} and regional and international initiatives to develop QKD networks \cite{sasaki_field_2011, wonfor_quantum_2021, eu_euroqci_2024, martin_madqci_2024, bersin_development_2024, chen_implementation_2025}.
Like WR, QKD systems primarily operate over optical fibre channels.
In their highest-speed implementations \cite{takesue_10ghz_2006,dixon_gigahertz_2008,boaron_simple_2018}, they require sub-nanosecond synchronization between nodes separated by tens to hundreds of kilometres, with jitter of the order of tens of picoseconds.
Similar requirements also apply to achieve high-quality entanglement distribution.

Today, many QKD deployments implement synchronization using custom, vendor-specific techniques, often optimized for a single point-to-point link.
Whilst effective for specific deployments, these approaches do not scale well to the large and heterogeneous networks envisioned for a future quantum internet.
The scale of deployed QKD networks is rapidly expanding, with a milestone 10,000 km QKD network deployed in China \cite{chen_implementation_2025}.
As this growth accelerates, a solid solution for shared timing and interoperability between equipment from multiple vendors will become essential \cite{etsi_qkd}.
Relying on WR for synchronization provides a standardized, vendor-agnostic timing layer, distributed alongside quantum and classical signals over the same fibre infrastructure. This would eliminate the proliferation of incompatible timing domains and significantly reduce overhead in multi-node QKD networks. 
Furthermore, WR unifies timing across the entire network, providing a common reference clock to all QKD systems, and its open and mature ecosystem aligns with ongoing standardization efforts in QKD \cite{etsi_qkd}, supporting interoperability and efficient resource usage as commercial quantum communication infrastructure continues to grow.

To date, WR has been demonstrated alongside entanglement distribution \cite{gerrits_white_2022, alshowkan_advanced_2022, schatz_practical_2023, rahmouni_100km_2024}, leading to significant improvement in observed fidelity compared to GPS \cite{alshowkan_reconfigurable_2021}, and has already been utilised to support QKD \cite{clivati_coherent_2022,magliano_twinfield_2025}.
Previous long-distance demonstrations of WR over links of several hundred kilometres have relied on cascaded amplification, with erbium-doped fibre amplifiers (EDFAs) placed at intervals of no more than \SI{80}-\SI{140}{\kilo\meter}~\cite{kaur_500km_2022, dierikx_white_2016}.
In certain channels, however, intermediate amplification is unavailable except directly at the end points, before and after the channel. A prominent example is long-haul submarine fibre used for low-latency financial connectivity, which typically consists of a single unrepeated span. These links naturally lend themselves to quantum applications, which do not admit the use of standard optical repeaters or amplifiers along the line. Indeed, some of these underwater long-haul fibre links has recently been used to demonstrate suitability for entanglement distribution and QKD~\cite{amies-king_quantum_2023, ribezzo_quantum_2023}. Similar constraints also occur in dark-fibre networks for metrology, regional terrestrial backbones, and other long spans where mid-link access is not feasible. These scenarios motivate the need to extend the reach of unrepeated WR, whilst maintaining the integrity of its signals.

In this work, we extend the reach of unrepeated WR from the previously demonstrated tens of kilometres to 300 km, constituting, to our knowledge, the longest such demonstration of WR to date.
Achieving this distance is a critical milestone toward deploying WR in real-world QKD networks, where nodes may be separated by hundreds of kilometres.
Additionally, we demonstrate that WR can operate in a configuration presenting significant asymmetry. This is a natural scenario occurring when two largely different wavelengths are used in the same optical fibre, or when the same wavelength is used in two different fibres of unequal length, one acting as the forward path and the other as the return path. This can happen in practice when the two directions are routed through different duct segments or carrier infrastructures, when restoration or protection paths differ significantly in length, or in long-haul links where terrestrial backhaul to the two endpoints follows different physical routes. Rather than being less general than a symmetric \SI{300}{\kilo\meter} demonstration, this is in fact more significant, as WR is designed to operate with only the small asymmetries arising from wavelength-dependent group velocity in optical fibre. Our result therefore demonstrates resilience against practical constraints in deployed environments, and the more straightforward symmetric \SI{300}{\kilo\meter} configuration would not present any additional technical challenge.

\section{Methods}

\begin{figure}[t]

  \begin{minipage}{0.53\textwidth}
    \begin{overpic}[width=\linewidth]{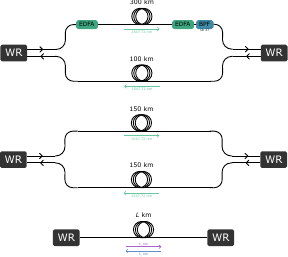}
    \put(0,82.5){\subcaptiontext*[1]{}}
    \phantomsubcaption\label{fig:300km_setup}
    \put(0,45){\subcaptiontext*[2]{}}
    \phantomsubcaption\label{fig:200km_setup}
    \put(0,9){\subcaptiontext*[3]{}}
    \phantomsubcaption\label{fig:7km_setup}
    \end{overpic}
  \end{minipage}\hfill%
  \begin{minipage}{0.44\textwidth}
    \begin{overpic}[width=\linewidth]{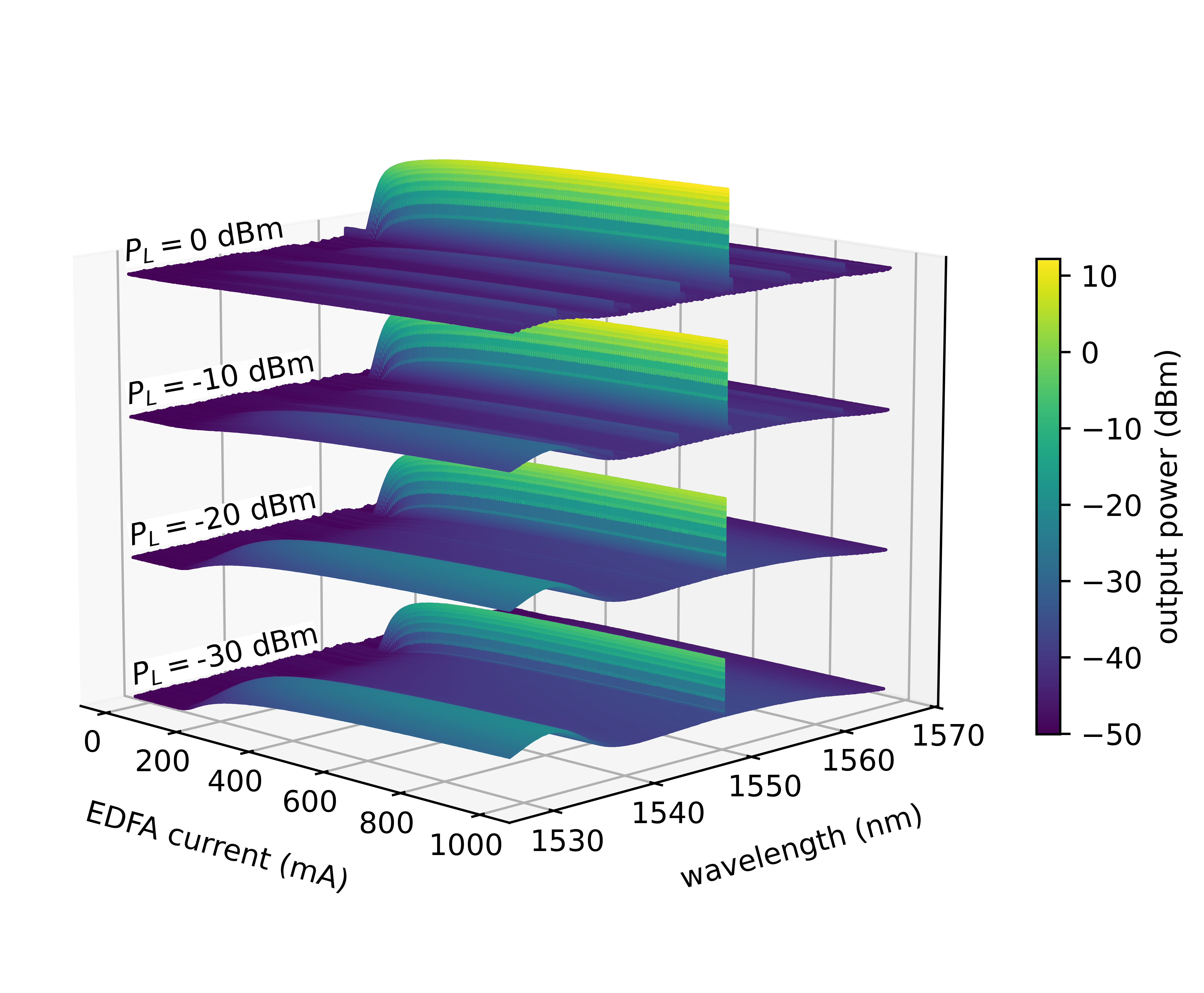}
    \put(2,68){\subcaptiontext*[4]{}}
    \phantomsubcaption\label{fig:edfa}
    \end{overpic}
    \begin{overpic}[width=\linewidth]{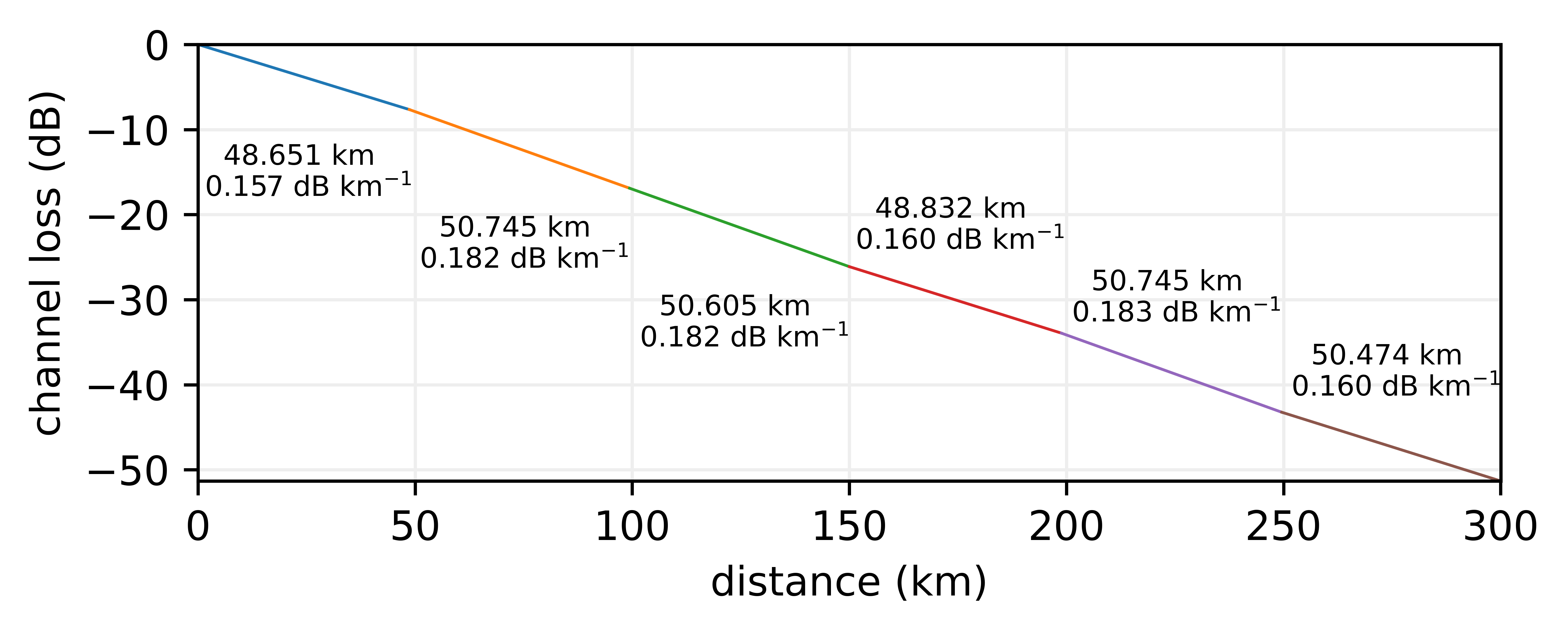}
    \put(2,40){\subcaptiontext*[5]{}}
    \phantomsubcaption\label{fig:channel_profile}
    \end{overpic}
  \end{minipage}
  \caption{
    Schematic diagrams of the three White Rabbit (WR) configurations considered in this work, and associated characterisation.
    \textbf{a} \textit{Asymmetric}: \SI{300}{\kilo\meter} duplex fibre (one arm of \SI{300}{\kilo\meter} and one of \SI{100}{\kilo\meter}) with one-way loss of \SI{\loss}{\deci\bel} for the \SI{300}{\kilo\meter} link.
    A single wavelength of \SI{1547.72}{\nano\meter} is used on both channels.
    \textbf{b}  \textit{Symmetric}: \SI{150}{\kilo\meter} duplex fibre, with a single wavelength.
    \textbf{c} \textit{Typical}: WR over $L$ \si{\kilo\meter} simplex fibre, with two wavelengths corresponding to the two directions of propagation.
    \textbf{d} Characterisation of EDFA gain and noise spectrum for a \SI{1550}{\nano\meter} \SI{1}{\kilo\hertz} linewidth laser input with optical power $P_L$. 
    \textbf{e} A representation of the six \SI{50}{\kilo\meter} stretches comprising the \SI{300}{\kilo\meter} channel, with their measured lengths and respective average loss coefficients.
  }
    \label{fig:experimental_setup}
\end{figure}

In our demonstration, we utilise two WR-LEN devices \cite{safran_wr-len_2024}, which implement the WR PTP core, with off-the-shelf small form-factor pluggable (SFP) transceivers.
WR-LEN devices provide two outputs under conventional master/slave operation: a \SI{10}{\mega\hertz} clock, and a \SI{20}{\milli\second} pulse-per-second signal (PPS).
Using a time-to-digital converter (TDC), we record the time of each PPS with picosecond precision in order to monitor the clock skew of the two WR devices over time.

The recommended configuration for the WR-LEN devices is using simplex fibre bi-directional (BiDi) 1000BASE-BX10-U SFPs with wavelengths \SI{1310}{\nano\meter}/\SI{1490}{\nano\meter}, which propagate in opposite directions in the same fibre.
However, WR has previously been demonstrated in duplex fibre \cite{dierikx_white_2016}, to enable bidirectional optical signals at \SI{1550}{\nano\meter} to minimise channel loss.
We adopt this approach to overcome the total loss of the link. More specifically, we utilise duplex fibre DWDM 10GBASE-ZR SFP+ transceivers at \SI{1547.72}{\nano\meter} (ITU grid Ch 37), with maximum launch power \SI{5}{\dBm} (\SI{3.2}{\milli\watt}) and sensitivity \SI{-23}{\dBm} (\SI{5}{\micro\watt}) giving a \SI{28}{\deci\bel} loss budget without amplification.
We introduce an erbium-doped fibre amplifier (EDFA) as a booster amplifier to increase the power launched into the fibre, and another EDFA as a pre-amplifier to increase the signal power to a level that meets the limited detector sensitivity of the SFP transceiver.
Both EDFAs nominally have a noise figure (NF) of less than \SI{5}{\decibel}, a small signal gain of around \SI{30}{\decibel} and a maximum output power of \SI{20}{\dBm} (\SI{100}{\milli\watt}).
The pump current of the EDFAs can be adjusted to increase the EDFA gain at the expense of increased noise.
This can clearly be seen in \Figure{edfa}, where we plot the EDFA output power in the range \SI{1530}{\nano\meter}--\SI{1570}{\nano\meter} measured with an optical spectrum analyser (OSA), over the full range of EDFA current, for increasing input laser power $P_L$.
In particular, we note the increasing EDFA noise with decreasing $P_L$ and increasing EDFA current.
For a successful connection between WR-LEN devices over a \SI{300}{\kilo\meter} fibre, the EDFA pump current must be chosen carefully to maintain the extinction ratio (ER) of the optical signal.

The PPS signal from the two WR-LEN devices is connected to a time-to-digital converter (TDC) which records the timestamps of the PPS pulses with picosecond precision and \SI{3}{\pico\second} RMS jitter.
We calculate the time error series $x_i$ from these timestamps, by finding the difference between the rising edge of the primary device PPS and the secondary device PPS.
When the WR-PTP is active, we expect one pulse per second from each device.
If there is a longer delay between consecutive events on a TDC channel, this indicates an instantaneous drop-out in the connection, and resulting failure of the WR-PTP.
When this occurs, the secondary device does not emit a PPS signal until it has successfully renewed the lock to the primary device.
As such, we discard the timestamps where there is no PPS from the secondary device, as we know this corresponds to a momentary drop-out in the connection.
Over the course of \SI{20}{\hour}, six drop-outs occur which result in a failure of the WR-PTP, and we observe that the WR-PTP resumes within \drouputDuration in each case.

Our channel consists of a mixture of ultra-low loss (ULL) and standard single-mode optical fibre, with a total length of \distance and total loss of \SI{\loss}{\deci\bel}, giving an average loss coefficient of \SI{0.17}{\deci\bel\per\kilo\meter}.
A loss profile of the channel is shown in \Figure{channel_profile}, which we determine from measurements of the individual spools' length and total loss.
We measure the length of an individual spool down to the metre level using a commercial optical time-domain reflectometer (OTDR).
To determine the spool loss, we use a telecoms-standard laser attenuated to the microwatt level to avoid any non-linear effects of the fibre.
Using two calibrated power meters, we simultaneously monitor the input and output power in order to eliminate the impact of laser power fluctuations, for \SI{100}{\second} with a measurement frequency of \SI{10}{\hertz}.
We then average over these 1000 measurements to determine the total loss of the fibre, and calculate the loss coefficient using the length measurement.
In general, our loss coefficient is lower than conventional deployed fibre, which tends to have higher loss due to routing via interconnects and stretches of legacy fibre with high loss.
Nevertheless, it is feasible to find a similarly low loss coefficient in the long-haul stretches of uninterrupted fibre which are the focus of this work \cite{amies-king_quantum_2023}.

The output power from the SFP transceiver is \SI{2.07}{\dBm}, which is amplified by the first EDFA to a fibre launch power of \SI{22.31}{\dBm}.
After the \SI{\loss}{\deci\bel} loss of the channel, \SI{-37.12}{\dBm} enters the second EDFA which outputs \SI{17.73}{\dBm} consisting mainly of a broad noise spectrum as shown in \Figure{edfa}.
An optical bandpass filter greatly attenuates the EDFA noise spectrum, before the signal is detected by the SFP transceiver.

In addition to the \SI{300}{\kilo\meter} result, we include three other WR configurations: \SI{150}{\kilo\meter} duplex as shown in \Figure{200km_setup}, to benchmark against a symmetric long-haul duplex system, and both \SI{150}{\kilo\meter} and \SI{7}{\kilo\meter} simplex as shown in \Figure{7km_setup} to represent the `ideal' performance of the WR devices in a conventional deployment at different length scales.
The \SI{150}{\kilo\meter} duplex demonstration utilises the same DWDM SFPs as the \SI{300}{\kilo\meter} setup, without any amplification to overcome the channel loss in either direction.
Accordingly, we find that no additional filtering is required as there is no EDFA to introduce a broad noise spectrum.
The \SI{150}{\kilo\meter} simplex demonstration is performed with long-range BiDi SFPs rated for \SI{160}{\kilo\meter}, with wavelengths \SI{1490}{\nano\meter} and \SI{1550}{\nano\meter}.
Finally, the \SI{7}{\kilo\meter} simplex demonstration uses the WR-prescribezribed BiDi SFPs, which use counter-propagating \SI{1310}{\nano\meter} and \SI{1490}{\nano\meter} signals in a common fibre up to \SI{10}{\kilo\meter} in length.

In the typical simplex fibre operation of WR, a small asymmetry may arise due to difference in propagation times of the two signal wavelengths.
This can be compensated by tuning the WR-LEN devices' asymmetry parameter $\alpha$, given by 

\[ \alpha = \frac{\delta_{MS}}{\delta_{SM}} - 1 \]

\noindent%
where $\delta_{MS},\delta_{SM}$ are the latencies due to fibre propagation in each direction between the leader (M) and follower (S) devices.
This is then converted to a natural number $\alpha_N$ for use with the WR PTP core as per the WR Calibration procedure \cite{daniluk_wr-calibration_2015}:

\[ \alpha_N = 2^{40} \left(\frac{\alpha - 1}{\alpha-2} - \frac{1}{2} \right)\]

\noindent%
which is stored as a signed 32-bit integer, corresponding to an approximate range of \num{-7.8e-3} to \num{7.8e-3} for $\alpha$.

For bidirectional configurations with simplex fibre, a given $\alpha$ parameter is valid for a particular combination of wavelengths and so can be calibrated in a laboratory prior to deployment.
In the event that pre-calibration is not feasible, the WR calibration guide outlines a process for in-situ calibration \cite{daniluk_wr-calibration_2015}.
In the dual fibre case, $\alpha$ must be calibrated depending on the specific length asymmetry of the fibre pair, which will be unique to that particular fibre pair.
Typically this asymmetry is of the order of metres and can easily be compensated by setting $\alpha$, and replicating in the two links the components shown in \Figure{300km_setup} for the \SI{300}{\kilo\meter} link.
However, in this work, we test a more challenging situation where the asymmetry is pushed to the level of \SI{200}{\kilo\meter}.
Whilst this abnormally large asymmetry would not be present in a typical deployment of WR, we show that the WR devices still achieve an excellent level of timing precision.

\section{Results}

We observe successful operation of WR time distribution over a link consisting of 300 km of uninterrupted optical fibre over almost 20 hours of operation.
In that time, WR dropped out six times; in each case, it re-established itself within \drouputDuration without any user intervention giving an uptime of \SI{99.86}{\percent}.

\begin{figure}
\centering
\begin{overpic}[width=0.49\linewidth]{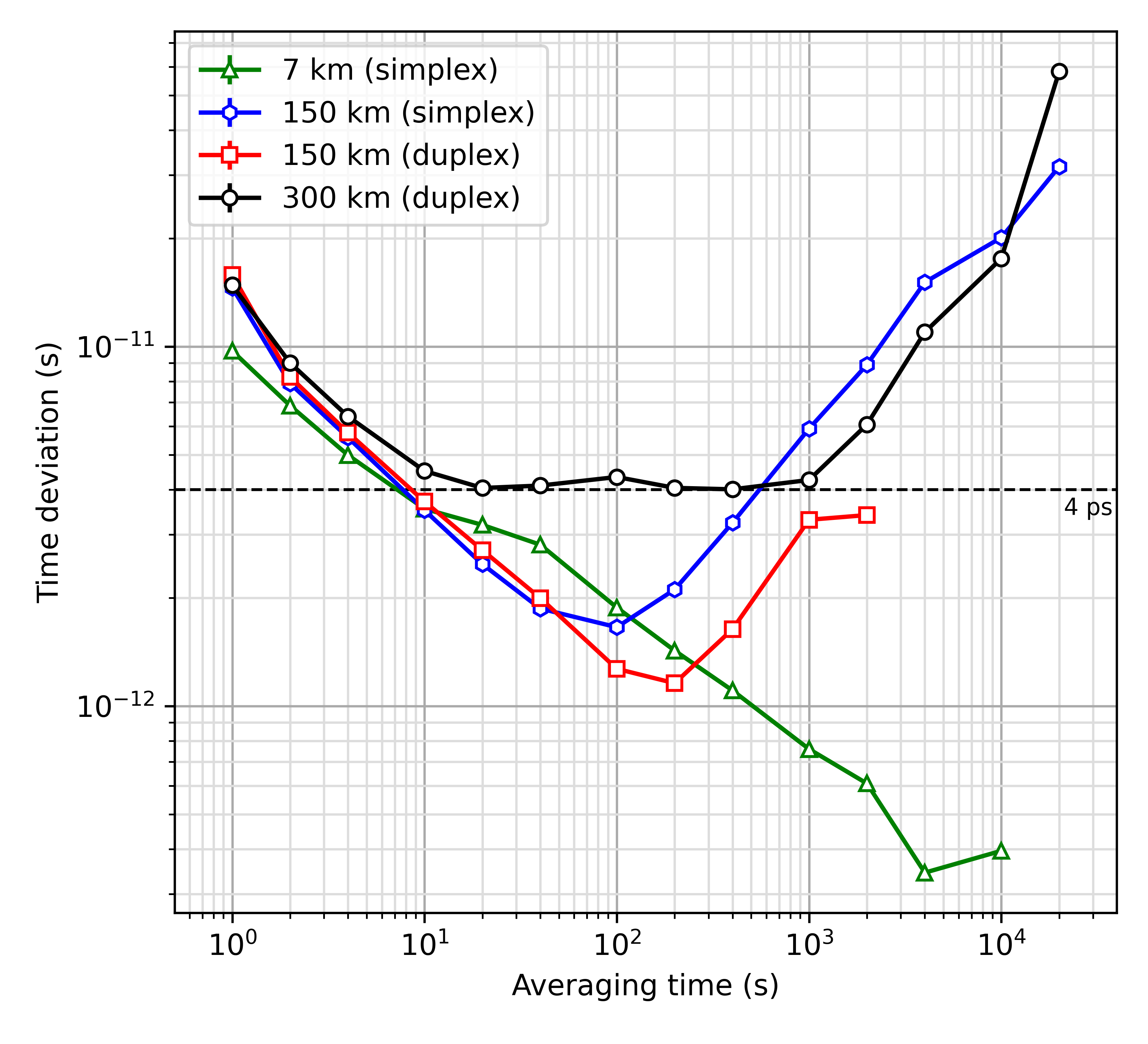}
\put(10,90){\subcaptiontext*[1]{}}
\phantomsubcaption\label{fig:tdev}
\end{overpic}
\begin{overpic}[width=0.49\linewidth]{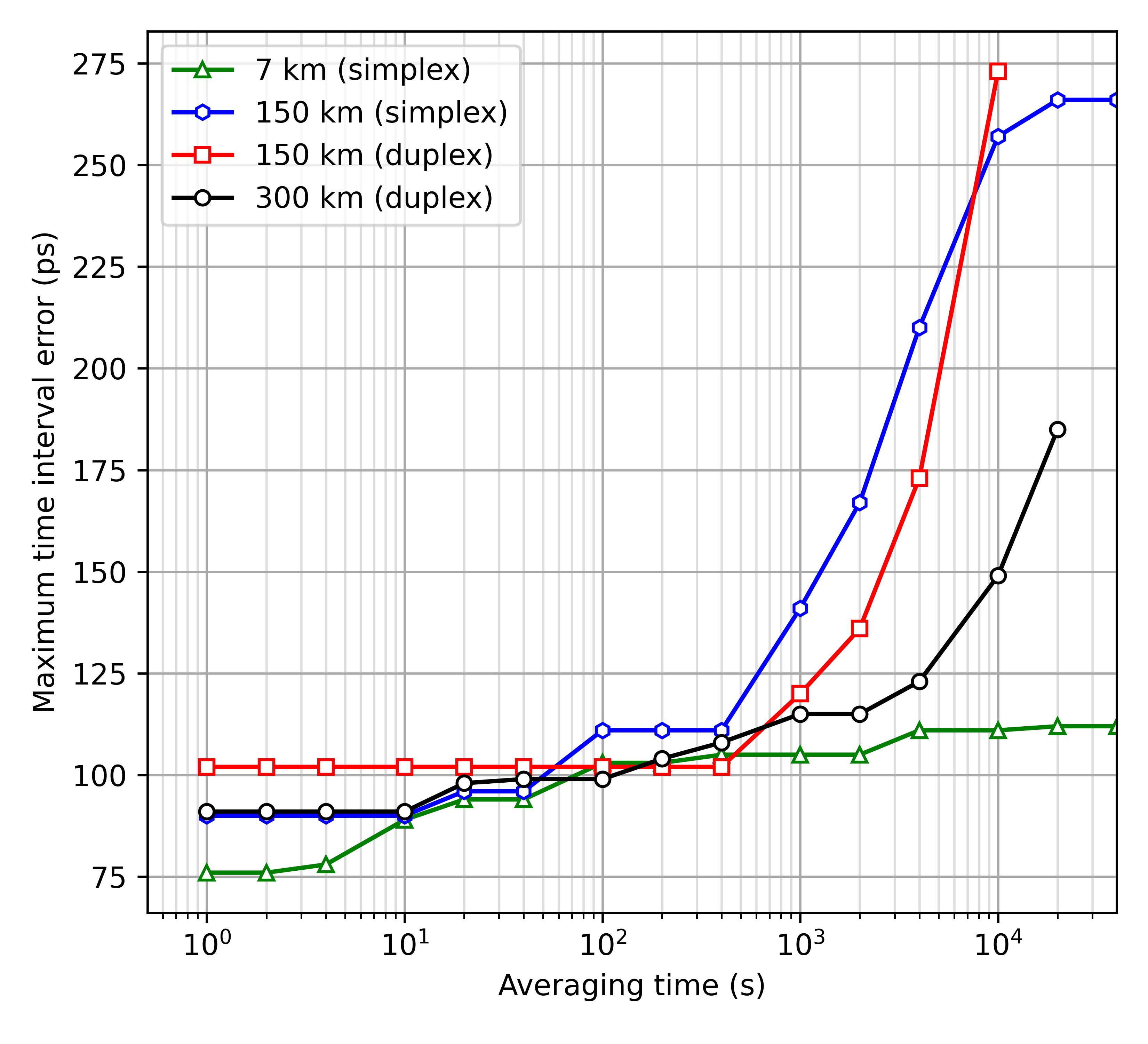}
\put(8,90){\subcaptiontext*[2]{}}
\phantomsubcaption\label{fig:mtie}
\end{overpic}
\caption{
  \textbf{a} Time deviation $\sigma_x (\tau)$ (\Equation{tdev}) and \textbf{b} maximum time interval error $\textrm{MTIE}(\tau)$ (\Equation{mtie}) of the the duplex \SI{300}{\kilo\meter} (black crosses), duplex \SI{150}{\kilo\meter} (red squares), simplex \SI{150}{\kilo\meter} (blue hexagons), and simplex \SI{7}{\kilo\meter} (green triangles) demonstrations.
}
\end{figure}

\Figure{tdev} shows the Allan time deviation $\sigma_x (\tau)$ of the two clocks for each of the three channel configurations considered, defined as \cite{allan_new_1990,allan_frequencydomain_1991, riley_handbook_2008}:
\begin{equation}\label{eq:tdev}
    \sigma_x^2 (\tau) = \frac{1}{6m^2(N-3m+1)}\sum_{j=1}^{N-3m+1}\left\{\sum_{i=j}^{j+m-1}\left[x_{i+2m} - 2x_{i+m} + x_i\right]\right\}^2
\end{equation}

\noindent where the $x_i$ are the $N$ phase error measurements in units of time, and $\tau = m\tau_0$ the averaging time with $\tau_0$ the interval between measurements.

For the \SI{300}{\kilo\meter} demonstration, a minimum time deviation of \SI{4.00(2)}{\pico\second} is achieved at averaging time $\tau = \SI{400}{\second}$.
In the context of QKD, this constitutes an exceptionally low timing uncertainty.
High-rate QKD demonstrations have been performed with detector jitter of the order of tens of picoseconds \cite{takesue_10ghz_2006,dixon_gigahertz_2008,boaron_simple_2018}, whilst state-of-the-art superconducting single-photon detectors achieve timing jitter of a few picoseconds \cite{korzh_demonstration_2020}, as do modern time-to-digital converters \cite{altruda_picotdc_2023}.

In addition, minimum time deviations of \SI{1.16(1)}{\pico\second} at $\tau = \SI{200}{\second}$ for \SI{150}{\kilo\meter} (duplex), \SI{1.69(1)}{\pico\second} at $\tau = \SI{100}{\second}$ for \SI{150}{\kilo\meter} (simplex), and \SI{344(2)}{\femto\second} at $\tau = \SI{4000}{\second}$ for \SI{7}{\kilo\meter} are observed.
In \Figure{tdev} we see that for \SI{300}{\kilo\meter} the time deviation plateaus at \SI{4}{\pico\second} between \SI{20}{\second} and \SI{1000}{\second} averaging times, due to an additional $\tau^0$ contribution corresponding to flicker phase noise \cite{allan_frequencydomain_1991}.
This is a noticeable difference from the \SI{150}{\kilo\meter} time deviation, which shows a clear minimum in that region.
Whilst the two demonstrations differ in total fibre length by \SI{100}{\kilo\meter}, in the \SI{300}{\kilo\meter} demonstration there is also a significant asymmetry of \SI{200}{\kilo\meter}.
Our results do not conclusively show whether this difference is due to the asymmetry or the greater channel length, however we expect this noise to be less pronounced in a symmetric \SI{300}{\kilo\meter} demonstration.

The \SI{7}{\kilo\meter} demonstration does not appear to show the limit of WR performance under conventional use.
Overall, the downwards trend of time deviation remains relatively linear with increasing averaging time $\tau$.
A measurement of longer duration would be required to determine where the time deviation reaches a limit.
Somewhat surprisingly, we cannot conclusively say that simplex fibre WR offers better performance than duplex fibre.
In fact, comparison of the \SI{150}{\kilo\meter} simplex and duplex results indicates the opposite although this may be attributable to environmental differences between the two measurements.
Further work is required to exhaustively explore the trade-off between simplex and duplex fibre for WR performance.
Whilst duplex fibre clearly requires double the channels per deployment, the separation of propagation directions means that both signals can exploit a single wavelength chosen for maximum transmission.

In \Figure{mtie}, we plot the maximum time interval error (MTIE) of the three demonstrations, given by \cite{riley_handbook_2008, bregni_measurement_1996, bregni_clock_1997}:

\begin{equation}\label{eq:mtie}
\textrm{MTIE}(\tau) = \max_{1\leq k \leq N-n} \left\{\max_{k\leq i\leq k+n} (x_i) - \min_{k\leq i\leq k+n} (x_i)\right\}
\end{equation}

\noindent which is a measure of absolute worst performance for a given time interval $\tau = n\tau_0$, where again the $x_i$ are the $N$ phase error measurements in units of time.

During the longest continuous period without a dropout, the maximum time interval error for the \SI{300}{\kilo\meter} demonstration remained below \SI{100}{\pico\second} for averaging times up to and including \SI{100}{\second}.
Similarly, the maximum time interval error for the \SI{150}{\kilo\meter} and \SI{7}{\kilo\meter} simplex demonstrations remained low before starting to increase around $\tau = \SI{100}{\second}$.
The \SI{150}{\kilo\meter} duplex demonstration shows slightly different behaviour, with an initial maximum time interval error of slightly over \SI{100}{\pico\second} that remains unexceeded until an averaging time of \SI{400}{\second} is reached.
As this is higher than the other results and remains constant for a large range of averaging times, it indicates a single error event which may not be truly representative of the typical WR performance that can be expected in that configuration.

Due to the restricted range of $\alpha_N$ supported by the WR PTP core, it is not possible to align the two clocks with such a large asymmetry.
For a \SI{300}{\kilo\meter} link, our asymmetry of \SI{200}{\kilo\meter} is almost two orders of magnitude greater than the supported maximum, resulting in a large skew observed between the two clocks.
If known, this skew can be accounted for experimentally as long as it remains relatively stable and, as we show here, WR retains precision of the order of picoseconds.

\section{Discussion}
\raggedbottom

In this work, we have demonstrated the feasibility of long-haul WR without intermediate amplification or repeaters, and related its performance to high-rate QKD and state-of-the-art single photon detection hardware.
We have also observed for the first time that WR can operate in highly asymmetric configurations, and that it is sufficiently resilient for use in challenging real-world fibre networks where a co-propagating fibre pair may be unavailable.
Notably, we have shown that the performance of WR over 300~km is sufficient to meet the timing requirements of high-speed applications over long stretches of single-span optical fibre, including high-rate QKD of \SI{1}{\giga\hertz} and above.
A time deviation on the order of picoseconds and maximum time interval error of less than \SI{100}{\pico\second} over \SI{100}{\second} indicates that negligible timing induced error would be introduced in QKD systems with repetition rates of several gigahertz, showing that even long-haul WR is ready to support the next generation of QKD systems.

These results also position WR as an ideal candidate for timing dissemination in entanglement distribution networks, where more precise timing leads to better visibility and better violation of Bell inequalities over distance \cite{alshowkan_advanced_2022}.
Entanglement distribution is particularly well-positioned to exploit the synchronised network architecture that WR inherently establishes, through many-user QKD networks exploiting wavelength demultiplexing of broadband entangled photon generation \cite{wengerowsky_entanglementbased_2018,joshi_trusted_2020,appas_flexible_2021,lingaraju_adaptive_2021,wang_dynamic_2022,clark_entanglement_2023}.

Furthermore, the low maximum time interval error demonstrates that long-range WR can enable sub-nanosecond timestamps for high-frequency financial transactions where correct ordering of operations is essential, particularly via unrepeated low-latency direct links.

Finally, applications where timing is critical are inherently vulnerable to attacks on the timing dissemination system itself.
This work demonstrates that quantum-secured WR over a typical QKD distance is achievable without intermediate amplification.
Indeed, a recent demonstration showed WR with QKD-based encryption of timing information over \SI{13}{\kilo\meter}, paving the way for an information-theoretically secure time-frequency transfer service \cite{meda_qkd_2025}.

To improve on these results, there are two key signal requirements of the SFP transceivers that could be further optimised. 
Firstly, the signal-to-noise ratio would potentially benefit from further filtering, and real-time tuning of EDFA gain parameters.
This would also allow optimisation of the launch power, to balance surplus loss from high power signals with sufficient power for effective booster amplification.

Another significant source of performance gains, albeit at the expense of deviating from off-the-shelf hardware, is to exploit more sensitive detectors.
SFPs typically have a sensitivity in the region of \SI{-30}{\dBm}, whilst classical detectors with sensitivity of \SI{-60}{\dBm} are readily available.
This, however, comes with the caveat that increasing sensitivity is accompanied by decreasing bandwidth; WR operates with a \SI{1000}{\mega\bit\per\second} optical link, and so detector bandwidth must be sufficient to resolve the requisite symbol rate.
This would also potentially enable lower launch powers in less lossy channels, minimising crosstalk between wavelength-division multiplexed classical and quantum signals.

For this work in particular, more detailed monitoring of the WR devices could help to explain the origin of the observed drop-outs.
They are relatively few and not obviously correlated to any change in the environmental conditions, making it challenging to determine their cause.
It is however clearly related to operating the SFPs at the very limit of their sensitivity, which could be better characterised with a suitable experiment.

Whilst WR is intended to operate with simplex fibre, our work indicates that the performance implications of duplex fibre are not necessarily clear-cut, as the duplex \SI{150}{\kilo\meter} result appears to outperform the 150 km simplex result.
We attribute this to a higher signal-to-noise ratio in the duplex case, as both directions benefit from the minimised loss coefficient at \SI{1550}{\nano\meter}.
Suitable amplification in the \SI{150}{\kilo\meter} simplex configuration could significantly improve the performance whilst simultaneously extending the achievable distance.

Further research into the feasibility of long-range WR over simplex fibre could lead to practical solutions for time-sensitive applications requiring strict synchronisation constraints.
In parallel, the coexistence of WR and QKD signals offers a promising route toward integrated quantum-classical infrastructure, particularly in long-distance deployments. 
Addressing the challenge of optimal wavelength channel assignment in such hybrid networks builds on a growing body of work exploring resource allocation in mixed quantum and classical optical systems \cite{patel_coexistence_2012,kumar_coexistence_2015,mao_integrating_2018,bahrani_wavelength_2018,bahrami_quantum_2020,dou_coexistence_2023,ruiz_routing_2025,wu_integration_2025}.

In addition to the experimental results, we make here the conceptual connection between WR, which is increasingly garnering interest as a commercially available off-the-shelf timing solution with excellent performance, and QKD, which is a mature quantum technology that has already seen industrial interest and strategically significant deployments.
The standardised nature of the WR protocol, off-the-shelf availability, open-source hardware and software, network-first architecture, and growing community positions it as the obvious choice for development of future QKD systems.
A number of recent developments to the WR technology could further consolidate this position, including support for higher data rates and elimination of the requirement for a hardware timing source which together could enable WR to provide the entire classical layer in a QKD system with minimal hardware overheads.

\begin{acknowledgments}
This work was supported by the Engineering and Physical Sciences Research Council Quantum Communications Hub (EP/T001011/1) and Integrated Quantum Networks Hub (EP/Z533208/1).
\end{acknowledgments}

\normalem
\bibliography{wr-300km}

\end{document}